\documentclass[MSNbibl,nameyear,dvips]{arxstspdf}
\usepackage{flushend}
\usepackage{stfloats}

\volume{26}
\issue{3}
\pubyear{2011}
\firstpage{317}
\lastpage{318}
\doi{10.1214/11-STS352A}
\referstodoi{10.1214/11-STS352}

\begin{document}
\begin{frontmatter}

\vspace*{9pt}
\title{Discussion of ``Is Bayes Posterior just Quick and Dirty Confidence?'' by D.~A.~S.~Fraser}
\runtitle{Discussion}
\pdftitle{Discussion of Is Bayes Posterior just Quick and Dirty Confidence? by D. A. S. Fraser}

\begin{aug}
\author[a]{\fnms{Christian P.} \snm{Robert}\corref{}\ead[label=e1]{xian@ceremade.dauphine.fr}}
\runauthor{C. P. Robert}

\affiliation{Universit\'e Paris Dauphine, IuF and CREST}

\address[a]{Christian P. Robert is Professor, CEREMADE,
Universit\'{e} Paris-Dauphine, 75775 Paris cedex 16, Senior Member of Institut Universitaire de
France and Senior Researcher, CREST, 92245 Malakoff cedex, France \printead{e1}.}
\end{aug}



\vspace*{-3pt}
\end{frontmatter}

While Don's paper does shed new insight on the evaluation of Bayesian
bounds in a frequentist light, the main
point of the paper seems to be a radical reexamination of the relevance
of the whole Bayesian approach to
confidence regions.
This is surprising given that the disagreement between classical and
frequentist perspectives is usually quite
limited (in contrast with tests) in that the coverage statements agree
to orders between $n^{-1/2}$ and
$n^{-1}$, following older results by Welch and Peers (\citeyear{WelPee63}).

First, the paper seems to contain a lot of apocryphal deeds attributed
to Thomas Bayes. My understanding of
the 1763 posthumous paper of Tho\-mas Bayes is one of a derivation of the
posterior distribution of a probability
parameter $\theta$ driving a~binomial observation $x\sim\mathcal
{B}(n,\theta)$. It thus fails to contain
anything about confidence statements or location parameters, in
relation with the ``translation invariance''
mentioned in the Introduction or in Section~7. As noted by Fienberg
(\citeyear{Fie06}), Thomas Bayes also does not introduce explicitly the constant
prior as a rule, even in his limited perspective, this had to wait for
Pierre Simon de Laplace twenty to thirty years later. Thanks to Don's
paper, however, I re-read Bayes' essay
(in Edward Deming's 1940 reprint) and found in both RULE~2 (page 400
and further) and RULE 3 (page 403 and
further) that Bayes approximated the (posterior) probability that the
parameter $\theta$ is between $x/n-z$
and $x/n+z$. However, a closer examination revealed that this part
(starting on page 399) had actually been\vadjust{\goodbreak}
written by Richard Price (even though Price mentions ``Mr Bayes's manuscript'').

My second and more important point of contention is that the Bayesian
perspective on confidence (or credible)
regions and statements does not claim ``correct coverage'' from a
frequentist viewpoint since it is articulated
in terms of the parameters. Probability calculus remains probability
calculus whether it applies to the
parameter space or to the observation space, making the comment about
\textit{the term probability
[being] less appropriate in the Bayesian weighted likelihood} quite
debatable. Following Jaynes (\citeyear{Jay03}),
``there is only one kind of probability.'' The title of the paper is
thus in complete contradiction with the
purpose of Bayesian inference and the chance identity occurring for
location parameters is a coincidence on
which one should not build sand castles.

I find Bayesian analysis neither quick (although it is logical) nor,
obviously, dirty (on the opposite, it proposes a more complete and more elegant inferential framework!).
Looking at a probability \mbox{evaluation} on the
parameter space being ``correct'' (Section 3) is also strange in that
the referential for a~Bayesian analysis is
the prior endowed space, not the conse\-quences on observable values that
have not been observed to paraphrase
Harold Jeffreys. A Bayesian~cre\-dible interval is therefore correct in
terms of the~pos\-terior distribution it is
derived from and it does~not address the completely different target of
finding a~fre\-quency-valid interval.
(The distinction made in the Bayesian literature, as, for example,
Berger, \citeyear{Ber85}, between confidence and credible
intervals is significant for those different purposes.) That a $\beta$
quanti\-le Bayesian confidence bound does
not exclude the true value of the parameter in $100\beta\%$ of the
observations is not a cause for worry when
considering only the observed $y_0$ and the example of Section~4 is
perfectly illustrating this perspective.
When I see on Figure 4(c) that the Bayesian coverage starts at~$1$
when $\theta=\theta_0$ I am indeed quite
happy with the fact that\vadjust{\goodbreak} this coherent procedure accounts for the fact
that $\theta$ cannot be lesser than
$\theta_0$. I thus strongly object to the dire conclusion of
\textit{Bayes
approach [being] viewed as a long
history of misdirection}! I also fail to understand what is the meaning
of ``reality'' in Section~7. When
running Bayesian inference, the parameter $\theta$ driving the observed
data is fixed but unknown. Having a
prior attached to it has nothing to do with ``reality,'' it is a
reference measure that is necessary for making
probability statements (or, quoting again from Jaynes, \citeyear{Jay03}, extending
the logics framework). Thus, the
apparently logical concern in Section~7 on \textit{how probabilities can
reasonably be attached to a constant} has
no raison d'\^etre, neither does the debate about \textit{where the prior
comes from} (Section 9). If the matter is
about improper versus proper priors (as hinted at by the comments about
marginalization paradoxes), this has
been extensively discussed in the literature and the difference seems
to me less important than the difference
between Bayes and generalized Bayes estimators.

While this is directly related to the above, the discussion in the
Paradigm section also confuses me.
Introducing a temporal order between $y_1$ and $y_2$ does not make
sense from a probabilistic viewpoint. Both
representations $f(y_1)f(y_2^0|y_1)$ and $f(y_2^0)f(y_1|y_2^0)$ are
equally valid. I note as a side remark that
the derivation of $f(y_1|y_2^0)$ as the collection of simulated $y_1$'s
for which $y_2=y_2^0$ is exactly the
starting point of the ABC (Approximate Bayesian calculation) algorithm
(Rubin, \citeyear{Rub84}; Pritchard et al., \citeyear{Prietal99}).

I further object to the debate about optimality (and the subsequent
relevance of Bayes procedures), as I~do believe
that decision theory brings a \mbox{useful} if formal representation of
statistical inference. The \textit{choice of the
criterion} that I understand as the choice of the loss function is
clearly important; however, it helps in
putting a meaning to notions like ``real'' or ``true'' or ``correct''
found in the paper. Changing the criterion
does change the outcome for the ``optimal'' interval, but the underlying
relevance of Bayesian procedures does
not go away. For instance, we proposed in Robert and Casella (\citeyear{RobCas94})
several of such losses for evaluating
confidence sets. The criticism found at the end of this Section 9 is
inappropriate in that the posterior
quantile is neither derived from a~loss function nor evaluated under a~specific loss function, since the
``nonzero'' curvature drawback stems from a frequentist perspective. Let
me also add that, even from\ a
frequentist perspective, strange and counter-intuitive phenomena can
occur, like the domination of the
classical confidence region by regions that are equal to the empty set
with positive probability (Hwang and
Chen, \citeyear{HwaChe86}).\looseness=1

In conclusion, I am quite sorry about the negati\-ve (and possibly
strident) tone of this discussion. However, I
do not see a convincing reason for opening afresh the Pandora box about
the (lack of) justifica\-tions for
the Bayesian approach, the true \mbox{nature} of probability and the
philosophical relevance of priors: The last
section is a nice and provocative enough collection of aphorisms,
although I doubt it will make a~dent in the
convictions of Bayesian readers. Baye\-sian credible intervals \textit{are
not} frequentist confidence intervals and
thus do not derive their optimality from providing an exact frequentist
coverage.



\begin{thebibliography}{9}

\bibitem[\protect\citeauthoryear{Berger}{1985}]{Ber85}
\begin{bbook}[mr]
\bauthor{\bsnm{Berger},~\bfnm{James~O.}\binits{J.~O.}}
(\byear{1985}).
\btitle{Statistical Decision Theory and {B}ayesian Analysis},
\bedition{2nd} ed.
\bpublisher{Springer}, \baddress{New York}.
\bid{mr={0804611}}
\bptok{imsref}%
\end{bbook}
\endbibitem

\bibitem[\protect\citeauthoryear{Fienberg}{2006}]{Fie06}
\begin{barticle}[mr]
\bauthor{\bsnm{Fienberg},~\bfnm{Stephen~E.}\binits{S.~E.}}
(\byear{2006}).
\btitle{When did {B}ayesian inference become ``{B}ayesian''?}
\bjournal{Bayesian Anal.}
\bvolume{1}
\bpages{1--40 (electronic)}.
\bid{mr={2227361}}
\bptok{imsref}%
\end{barticle}
\endbibitem

\bibitem[\protect\citeauthoryear{Hwang and Chen}{1986}]{HwaChe86}
\begin{barticle}[mr]
\bauthor{\bsnm{Hwang},~\bfnm{Jiunn~Tzon}\binits{J.~T.}} \AND
  \bauthor{\bsnm{Chen},~\bfnm{Jeesen}\binits{J.}}
(\byear{1986}).
\btitle{Improved confidence sets for the coefficients of a linear model with
  spherically symmetric errors}.
\bjournal{Ann. Statist.}
\bvolume{14}
\bpages{444--460}.
\bid{doi={10.1214/aos/1176349932}, issn={0090-5364}, mr={0840508}}
\bptok{imsref}%
\end{barticle}
\endbibitem

\bibitem[\protect\citeauthoryear{Jaynes}{2003}]{Jay03}
\begin{bbook}[mr]
\bauthor{\bsnm{Jaynes},~\bfnm{E.~T.}\binits{E.~T.}}
(\byear{2003}).
\btitle{Probability Theory}.
\bpublisher{Cambridge Univ. Press}, \baddress{Cambridge}.
\bid{doi={10.1017/CBO9780511790423}, mr={1992316}}
\bptok{imsref}%
\end{bbook}
\endbibitem

\bibitem[\protect\citeauthoryear{Pritchard et~al.}{1999}]{Prietal99}
\begin{barticle}[pbm]
\bauthor{\bsnm{Pritchard},~\bfnm{J.~K.}\binits{J.~K.}},
  \bauthor{\bsnm{Seielstad},~\bfnm{M.~T.}\binits{M.~T.}},
  \bauthor{\bsnm{Perez-Lezaun},~\bfnm{A.}\binits{A.}} \AND
  \bauthor{\bsnm{Feldman},~\bfnm{M.~W.}\binits{M.~W.}}
(\byear{1999}).
\btitle{Population growth of human Y chromosomes: A study of Y chromosome
  microsatellites}.
\bjournal{Mol. Biol. Evol.}
\bvolume{16}
\bpages{1791--1798}.
\bid{issn={0737-4038}, pmid={10605120}}
\bptok{imsref}%
\end{barticle}
\endbibitem

\bibitem[\protect\citeauthoryear{Robert and Casella}{1994}]{RobCas94}
\begin{barticle}[mr]
\bauthor{\bsnm{Robert},~\bfnm{Christian~P.}\binits{C.~P.}} \AND
  \bauthor{\bsnm{Casella},~\bfnm{George}\binits{G.}}
(\byear{1994}).
\btitle{Distance weighted losses for testing and confidence set evaluation}.
\bjournal{Test}
\bvolume{3}
\bpages{163--182}.
\bid{doi={10.1007/BF02562679}, issn={1133-0686}, mr={1293113}}
\bptok{imsref}%
\end{barticle}
\endbibitem

\bibitem[\protect\citeauthoryear{Rubin}{1984}]{Rub84}
\begin{barticle}[mr]
\bauthor{\bsnm{Rubin},~\bfnm{Donald~B.}\binits{D.~B.}}
(\byear{1984}).
\btitle{Bayesianly justifiable and relevant frequency calculations for the
  applied statistician}.
\bjournal{Ann. Statist.}
\bvolume{12}
\bpages{1151--1172}.
\bid{doi={10.1214/aos/1176346785}, issn={0090-5364}, mr={0760681}}
\bptok{imsref}%
\end{barticle}
\endbibitem

\bibitem[\protect\citeauthoryear{Welch and Peers}{1963}]{WelPee63}
\begin{barticle}[mr]
\bauthor{\bsnm{Welch},~\bfnm{B.~L.}\binits{B.~L.}} \AND
  \bauthor{\bsnm{Peers},~\bfnm{H.~W.}\binits{H.~W.}}
(\byear{1963}).
\btitle{On formulae for confidence points based on integrals of weighted
  likelihoods}.
\bjournal{J. Roy. Statist. Soc. Ser. B}
\bvolume{25}
\bpages{318--329}.
\bid{issn={0035-9246}, mr={0173309}}
\bptok{imsref}%
\end{barticle}
\endbibitem
\vspace*{-2pt}
\end{thebibliography}
\end{document}